\documentclass{iopart}

\usepackage{graphicx}
\usepackage{setstack,amssymb}

\begin{document}

\title[Maximum principle for sequence evolution]{Maximum principle 
and mutation thresholds for four-letter sequence evolution}

\author{T Garske and U Grimm}

\address{Applied Maths Department, Faculty of Mathematics and
Computing, The Open University, Walton Hall, Milton Keynes, MK7 6AA,
UK}

\ead{\mailto{t.garske@open.ac.uk},\mailto{u.g.grimm@open.ac.uk}}

\begin{abstract}
A four-state mutation--selection model for the evolution of
populations of DNA-sequences is investigated with particular interest
in the phenomenon of error thresholds.  The mutation model considered
is the Kimura 3ST mutation scheme, fitness functions, which determine
the selection process, come from the permutation-invariant class.
Error thresholds can be found for various fitness functions, the phase
diagrams are more interesting than for equivalent two-state
models. Results for (small) finite sequence lengths are compared with
those for infinite sequence length, obtained via a maximum principle
that is equivalent to the principle of minimal free energy in physics.
\end{abstract}


\section{Introduction}

In population genetics, one aims to model the evolution of a
population, taking into account evolutionary forces such as mutation,
selection, recombination, or migration, using either a stochastic
formulation, to incorporate the effects of a finite population, or a
deterministic formulation, to emulate the limit of infinite population
size. For a review, see \cite{BG00}.

In this paper, we are concerned with a deterministic
mutation--selection model for a haploid\footnote{Haploid organisms
have only one copy of each chromosome, and usually reproduce
asexually. Examples are viruses, bacteria and most fungi. In diploid
organisms, the genetic information exists in two copies, one stemming
from the father and one from the mother. } population (or a
\addtocounter{footnote}{-1} diploid\footnotemark\ population without
dominance). These types of models have been subject to investigation
in various different settings, see \cite{Bur00}. The questions one is
typically interested in comprise mutation--selection balance, mean
quantities and variances, and the mutation load, i.\ e., the
dependence of the mean fitness on the mutation rates.

One class of mutation--selection models are sequence space models,
where, inspired by the structure of the DNA, genotypes are modelled as
sequences written in a two- or four-letter alphabet, where each letter
stands for one of the purines adenine and guanine, or the pyrimidines
cytosine and thymine (or uracil in the case of RNA sequences).  The
majority of work in this field is done for two-state models, where one
state is identified with purines and the other with pyrimidines, or,
alternatively, one stands for a wildtype and the other for a mutant
site; results for models that incorporate the full for-letter
structure of the DNA are scarce \cite{HWB01, GG04}.

The advantage of sequence space models is that the mutation process is
straightforward to model. A drawback is, however, that the modelling
of the selection process is less clear. Selection comes in as
different reproduction rates that are assigned to different
genotypes. In nature, these are influenced by the phenotype, and the
mapping from genotypes to phenotypes is highly complex, such that any
feasible modelling of selection is necessarily simplistic.

Due to the linear structure and limited alphabet of the DNA, sequence
space models are very similar to some physical models, such as
one-dimensional Ising models or quantum spin chains. In fact, in an
appropriate formulation, they are equivalent to these \cite{Leu86,
BBW97}. The realization of this equivalence made the well developed
methods and tools from statistical physics accessible in the
population genetics context. The analogies between the physical and
the biological models are however involved, as the observables and
quantities of interest are slightly different in the different
disciplines. One example for the transfer of knowledge from physics to
biology is the derivation of a maximum principle for the population
mean fitness, that finds its analog in physics in the principle of
minimal free energy. This has been shown for a two-state model in
\cite{HRWB02}, generalised to a four-state model in \cite{GG04}, and
derived for a very general class of evolution models in biology in
\cite{BBBK}.

Of particular interest is the phenomenon of
a mutation driven {\em error threshold}. This phenomenon is based on
the antagonistic action of mutation and selection.  For low mutation
rates, selection dominates the population, such that its distribution
is going to be clustered around the optimal sequence and thus is well
localized in sequence space. At high mutation rates, on the other
hand, mutation outweighs the effect of selection and the population
distribution is an equidistribution in sequence space, which certainly
leads to extinction. If the transition from the localised to the
delocalized population distribution is very sharp and occurs in a very
narrow regime of mutation rates, one speaks of an error threshold.
This phenomenon is equivalent to phase transitions in physics, where
the temperature takes the role of the mutation rate.

The existence of error thresholds in biology has been predicted in the
framework of the quasi-species model \cite{Eig71}, and attracted
considerable interest ever since \cite{Leu87, Tar92, MS95, DCCC98,
HRWB02}. More recently, experimental indications for the existence of
error thresholds in virus population have been reported, see for
instance \cite{DH97, CCA01}.

Existence and properties of the error thresholds depend on the model
of selection. Most theoretical investigations of the error threshold
phenomenon are done for two-state models, the only work considering
four-state models we are aware of is \cite{HWB01}, and this is limited
to a single fitness function. The results indicate however more
involved phase diagrams than in the two-state case, warranting further
investigations.

The scope of this work is to investigate a four-state mutation--selection
model with particular attention for the phase diagrams with a number
of different fitness functions using the maximum principle for the
population mean fitness from \cite{GG04} to determine the error
thresholds.

The outline of the paper is as follows. In section 2, the general
setup of our deterministic mutation--selection model is stated,
introducing the quantities of interest. This is applied to the case of
a four-state sequence space model in section 3. Section 4 introduces
the major simplification of a permutation invariant fitness and
presents the resulting reduction of the type space. In section 5, we
recall the maximum principle from \cite{GG04}, which is applied to the
investigation of error thresholds in section 6. Finally, we close with
our conclusions.

\section{Mutation selection models}
The mathematical setup of a deterministic mutation--selection model
for a haploid population (or a diploid one without dominance) in
time-continuous formulation is as follows.  Individuals are identified
with genotypes, i.\ e., their fitness (or reproduction rate) is solely
determined by the genotype and thus environmental effects are
neglected. The number of genotypes is finite. In the case of sequences
of a fixed length $N$ written in a four-letter alphabet, there are
$4^N$ different sequences and thus genotypes. The set of genotypes
forms the the type space $\mathfrak{S}$, which is also called sequence
space in case of sequences as genotypes.

\subsection{Population}
The population at any time $t$ is described by the frequencies $p_i$
of each type $i\in\mathfrak{S}$, which are collected in the
$|\mathfrak{S}|$-dimensional vector $\bi{p}(t)$, where $|\mathfrak{S}|$
is the cardinality of the type space. The population is a
probability distribution and normalized such that $\sum_ip_i=1$.

\subsection{Evolution of the population}
Evolution is modelled in a time-continuous formulation.  Mutation and
selection are treated as two independent processes going on in
parallel.  In this case, the possible events are birth and death
events for each type $i$, and mutation events from type $j$ to $i$,
which happen with rates $b_i$, $d_i$ and $m_{ij}$, respectively. The
effective reproduction rate is given by the difference of birth and
death rates, $r_i=b_i-d_i$. The reproduction rates are collected in
the diagonal reproduction matrix $\mathcal{R}=(r_i)$.  Similarly, the
mutation rates are the entries of the mutation matrix
$\mathcal{M}=(m_{ij})$. The diagonal entries of the mutation matrix
are chosen such that $\mathcal{M}$ is a Markov generator and fulfils
the condition $\sum_i\mathcal{M}_{ij}=0$.  The time evolution operator
$\mathcal{H}$ is given as the sum of reproduction and mutation matrix
$\mathcal{H=R+M}$.

The evolution of the population is then given by the evolution
equation
\begin{equation}
\label{1}
\dot{\bi{p}}(t) =\left[\mathcal{H} -\bar{r}(t)\mathbf{1}\right]
\bi{p}(t) \;,
\end{equation}
where $\bar{r}(t)=\sum_ir_ip_i(t)$ is the population mean fitness, and
$\mathbf{1}$ is the unity matrix.

In equilibrium, $\dot{\bi{p}}=0$, thus this equation reduces to an
eigenvalue equation of $\mathcal{H}$.  All equilibrium quantities
shall be denoted by omitting the argument $t$. So $\bar{r}$ is the
population mean fitness in equilibrium for instance.  Assuming
irreducibility for $\mathcal{M}$, Perron-Frobenius theory \cite{Kar66}
applies, which implies that the leading eigenvalue $\bar{r}$ is
non-degenerate, and the corresponding (right) eigenvector $\bi{p}$ is
strictly positive, as required for a probability distribution.

\subsection{Relative reproductive success and ancestral distribution}
Similarly to the population distribution, there is another
distribution that will prove important in this model, namely the {\em
ancestral distribution}.

\begin{figure}
\begin{center}
\includegraphics{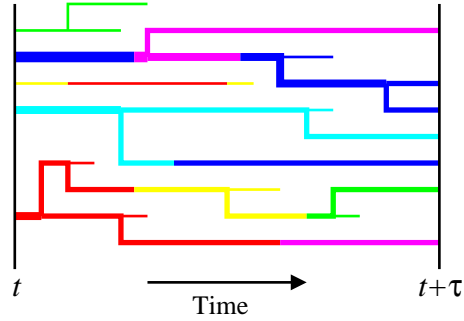}
\end{center}
\caption{\label{Fig1}Schematic population: lineages symbolize
individuals, colours indicate their types. Colour changes mark
mutation events, whereas branching lines represent births and ending
lines stand for deaths. The thickness of a line at any time denotes
the number of offspring of the individual at time $t+\tau$.}
\end{figure}
Figure \ref{Fig1} shows schematically a population evolving in time in
terms of a multitype branching process. The population distribution of
this population at any time (between $t$ and $t+\tau$) is easily
accessible by counting the number of lines of each type (colour) at
that time and normalizing it.

The relative reproductive success $z_i(\tau,t)$ of type $i$ at time
$t$ with respect to $t+\tau$ is proportional to the number of
offspring at time $t+\tau$ of all individuals of type $i$ at time $t$
divided by the number of type $i$ individuals at time $t$. As shown in
\cite{HRWB02}, for infinite populations in equilibrium
($t,\tau\rightarrow\infty$), the relative reproductive success
$\bi{z}$ is given by the left Perron-Frobenius eigenvector of $\mathcal{H}$.

The ancestral distribution $\bi{a}(\tau,t)$ is derived by tracing back
each line from $t+\tau$ to $t$ and counting it as the type it is at
$t$, irrespective of its type at $t+\tau$. As this is a probability
distribution, it has to be normalized such that $\sum_ia_i(\tau,t)=1$.
Thus we get $a_i(\tau,t)=z_i(\tau,t)p_i(t)$, with a suitable
normalization of $\bi{z}$, cf.\ \cite{HRWB02}, such that in
equilibrium the $a_i$ are given by both the left and right
PF-eigenvectors, $\bi{z}$ and $\bi{p}$.

\subsection{Ancestral and population averages}
Every quantity, say $f$, that is assigned to each type
$i\in\mathfrak{S}$ can be averaged with respect to both the ancestral
and population distribution. The population mean of $f$ is defined as
\begin{equation}
\label{2}
\bar{f}(t):= \sum_i f_ip_i(t)\;,
\end{equation}
the population mean fitness from equation (\ref{1}) is an important
example. The ancestral mean of $f$ is given by
\begin{equation}
\label{3}
\hat{f}(\tau,t):= \sum_i f_ia_i(\tau,t) \;.
\end{equation}

\section{The four-state model}
When modelling DNA or RNA evolution, the genotypes become sequences,
that are written in a four-letter alphabet $\Sigma$ (or binary
sequences in a simplified version). Here, we restrict ourselves to
sequences of fixed length $N$, and thus the sequence space
$\mathfrak{S}_N$ is the set of all four-state sequences $s$ of length
$N$, $\mathfrak{S}_N=\Sigma^N$ with $\Sigma=\{\tt A,G,C,T\}$ and
$|\mathfrak{S}_N|=4^N$.

\subsection{Mutation model}
\begin{figure}
\begin{center}
\includegraphics[width=3cm]{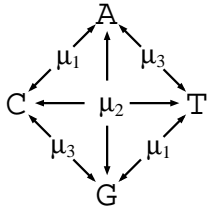}
\end{center}
\caption{\label{Fig2}Kimura 3ST mutation scheme.}
\end{figure}

The only mutation process taken into account is that of single point
mutations, disregarding more complex events such as multiple
mutations, deletions or insertions. This is known as the {\em single
step mutation model} \cite{OK73}. Site independent mutation rates are
assumed.  With four different nucleotides, there are 12 different
replacements of one by another. The rates for these are specified by
the {\em Kimura 3ST mutation scheme} \cite{Kim81,SOWH96}, cf.\ Figure
\ref{Fig2}, which uses only 3 different mutation rates, $\mu_1,\mu_2$
and $\mu_3$.

Common simplifications of this model are the {\em Jukes-Cantor
mutation scheme} \cite{JC69}, where all three mutation rates are the
same, $\mu_k:=\mu$, $k=1,2,3$ and the {\em Kimura 2 parameter model}
\cite{Kim80}, where $\mu_1\equiv\mu_3:=\mu$, which takes into account
only the different mutation rates for {\em transitions}, which are
replacements of one of the purines ({\tt A} or {\tt G}) by the other,
or of one of the pyrimidines ({\tt C} or {\tt T}) by the other, and
{\em transversions}, which are replacements of a purine by a pyrimidine
and vice versa. This is justified by the observation that transitions,
which occur at rate $\mu_2$, are found to have a far higher rate than
the two types of transversions, which happen at similar rates $\mu_1$
and $\mu_3$.

A convenient measure for closeness of two sequences is the Hamming
distance $d_H$ \cite{vLi82}, which counts the number of sites
at which two sequences $s,s'$ differ. In the single step mutation
model, this determines how many mutational steps away from each other
the two sequences are. Using the Kimura 3ST mutation scheme, one needs
a 3-dimensional Hamming distance $\bi{d}_H(s,s')$, to account for the
different types of mutations. The $k$-th entry of the Hamming
distance, $d_{H,\, k}$ simply counts the number of sites at which the
sequences $s$ and $s'$ differ in a way such that they can mutate into
each other with rate $\mu_k$.

The entries of the mutation matrix are given by
\begin{equation}
\label{3.1}
\mathcal{M}_{s,s'}=\cases{
\mu_1/N & for $\bi{d}_H(s,s')=(1,0,0)^T$\\
\mu_2/N & for $\bi{d}_H(s,s')=(0,1,0)^T$\\
\mu_3/N & for $\bi{d}_H(s,s')=(0,0,1)^T$\\
-(\mu_1+\mu_2+\mu_3) & for $s=s'$\\
0 & else \quad .
}
\end{equation}

\subsection{Representation of sequences}
Defining a reference type, which is usually (but not necessarily)
taken to be the {\em wildtype}, i.\ e., the type with maximal fitness,
we can conveniently change the alphabet $\Sigma$ from $\{{\tt
A,G,C,T}\}$ to $\{0,1,2,3\}$. This is done by assigning the zero
string to the reference type, $s_{\rm ref}=00\ldots0$. The
representation of any other sequence is obtained relative to the
reference type. Comparing a sequence $s$ with the wildtype $s_{\rm
ref}$, we assign $0$ to $s$ at each site where the two sequences
match; at sites where they differ, we assign $1$, $2$ or $3$,
depending on the type of mutation (determined by the respective
mutation rate).

Using this representation we can define the mutational distance
$\bi{d}$ as the 3-dimensional Hamming distance to the wildtype, 
\begin{equation}
\label{4}
\bi{d}= \left( \begin{array}{c} d_1 \\ d_2 \\ d_3 \end{array} \right)
= \left( \begin{array}{c} \#1 \\ \#2 \\ \#3 \end{array} \right)
\end{equation}
with $0\le d_1,d_2,d_3$ and the total mutational distance
$d=\sum_{k=1}^3d_k\le N$. The number of wildtype sites is given by
$d_0=N-\sum_{k=1}^3d_k$. The mutational distance determines how many
mutational steps of each type a sequence is away from the wildtype.

\section{Permutation invariant fitness model}
Whereas the mutation model is well justified on the microscopic level,
the picture is different for the reproduction rates. Only little is
known about the nature of realistic fitness functions, and any
realistic fitness function would be highly complex and probably too
hard to tackle analytically. Usually, the choice of fitness function
is determined rather by feasibility than by closeness to reality.

One rather common, though often implicitly used, assumption is that of
 a permutation-invariant fitness function. This means that the
 reproduction rate of a sequence depends only on its mutational
 distance to the wildtype, not on the order of the sequence and
 implies that all sequences with the same mutational distance have the
 same fitness.

Of course this assumption is highly unrealistic, as in nature the
location of a mutation (whether it lies in a coding or non-coding
region of the genome, for instance) will certainly influence the
fitness. Having said that, the accumulation of many mutations with
small effects on the fitness is described surprisingly well by a
permutation-invariant fitness.  In contrast to, for instance, magnetic
systems in physics, where the interactions are rather short ranged,
and therefore next-neighbour interactions seem more appropriate than
permutation-invariant ones, in the genome the interactions are
typically long-ranged, and therefore the permutation-invariant fitness
is a natural choice.

For a permutation-invariant fitness, we can reduce the dimensions
of the type space by a procedure that is known as lumping in the
theory of Markov chains (\cite{KS60}, Chapter VI). The essence here is
that multiple classes are lumped together in one class, such that a
new Markov process with fewer classes emerges. In our case we want to
lump together all sequences with the same fitness, such that the
lumped types are determined by the mutational distance $\bi{d}$. This
implies that the lumping is compatible with the mutation scheme, which
imposes the structure on the type space.

\subsection{The mutational distance space}
The sequence space $\mathfrak{S}$ contains all $\{0,1,2,3\}$-sequences
$s$ of length $N$, $\mathfrak{S}=\Sigma^N$, and thus
$|\mathfrak{S}|=4^N$. The neighbourhood or structure of this space is
determined by the mutation model: Two sequences are neighbours if and
only if they can mutate into one another with a single mutational
step, i.\ e.\, they differ only at a single site. Thus each sequence
has $3N$ neighbours as there are N sites and three types of mutation
at each site.

The lumped type space, or {\em mutational distance space}
$\mathcal{S}$ contains the $(N+1)(N+2)(N+3)/6$ different mutational
distances $\bi{d}$.  Going from the original sequence space
$\mathfrak{S}$ to the mutational distance space $\mathcal{S}$, the
neighbourhood is maintained, i.\ e., those and only those sequences
$s'$ that are neighbours of a particular sequence $s$ are in classes
$\bi{d}'$ that are neighbours of $\bi{d}$, the mutational distance of
$s$.  As $\mathcal{S}$ contains all possible mutational distances
$\bi{d}$, it can be visualized as a simplex in $\mathbb{N}^3$.

Describing the neighbourhood of the mutational distances, it is useful
to classify the sites of a sequence into wildtype and mutant sites,
where the wildtype sites are those denominated by $0$, whereas the
mutant sites are the $1,2,3$-sites. A mutation at a wildtype site
changes this site to a mutant site and thus increases the total
mutational distance $d$ to $d'=d+1$. This is symbolised by
\begin{equation}
\label{5}
\bi{d}\rightarrow\bi{d}'=\bi{d}+\bi{e}_k \quad\mbox{with}\quad 
\bi{e}_k=\cases{
(1,0,0)^T& for $k=1$ \\(0,1,0)^T& for $k=2$ \\(0,0,1)^T& for $k=3$ \;.
} 
\end{equation}
These mutations occur at rate $\mu_kd_0/N$. 

For a mutation at a mutant site, there are two possibilities. Firstly,
it can be a mutation from type $k$ back to the wildtype with rate
$\mu_k$ and thus decrease the total mutational distance; in this case
the mutational direction is given by $-\bi{e}_k$ with the $\bi{e}_k$
from equation (\ref{5}). Secondly, it can be a mutation to another
type of mutant site, which does not change the total mutational
distance $d$, but changes the components of the mutational distance
$\bi{d}$ to $\bi{d}'=\bi{d}\pm\bi{e}_{+k,-\ell}$ with
\begin{equation}
\label{6}
\bi{e}_{+k,-\ell}=\cases{
(1,-1,0)^T& for $k=1,\ell=2$\\
(1,0,-1)^T& for $k=1,\ell=3$\\
(0,1,-1)^T& for $k=2,\ell=3$
}\;.
\end{equation}
These mutations happen at rate $\mu_md_\ell/N$ and $\mu_md_k/N$, with
$\{k,\ell,m\}=\{1,2,3\}$, respectively.

\subsection{The lumped reproduction and mutation matrices}
As we assume that fitness is permutation invariant and we lump
together all sequences with the same mutational distance, the lumped
reproduction matrix $\bi{R}$ is diagonal, too, and it is easily
obtained as $\bi{R}_{\bi{d}}=\mathcal{R}_{s}$ with $\bi{d}=\bi{d}(s)$.

The lumped mutation matrix $\bi{M}$ differs more from the original
$\mathcal{M}$. The only nonzero entries (apart from the diagonal) are
the mutation rates between neighbouring mutational distances. They are
given by
\begin{equation}
\label{7}
\eqalign{
\bi{d}\rightarrow\bi{d}+\bi{e}_k: & \bi{M}_{\bi{d}+\bi{e}_k,\bi{d}}
=\mu_k d_0/N \\
\bi{d}\rightarrow\bi{d}-\bi{e}_k: & \bi{M}_{\bi{d}-\bi{e}_k,\bi{d}}
=\mu_k d_k /N  \\
\bi{d}\rightarrow\bi{d}+\bi{e}_k-\bi{e}_\ell:\quad & 
\bi{M}_{\bi{d}+\bi{e}_k-\bi{e}_\ell,\bi{d}}
=\mu_m d_\ell /N 
}
\end{equation}
with $\{k,\ell,m\}=\{1,2,3\}$, taking into account the number of
different sites at which the mutation can take place with the same
effect.  Using the same diagonal entries as in $\mathcal{M}$, i.\ e.,
\mbox{$M_{\bi{d,d}}=-\sum_k\mu_k$}, the mutation matrix is still a
Markov generator. Note that $\bi{M}$ is, however, not symmetric in
contrast to $\mathcal{M}$.  For details of the lumping procedure, see
\cite{KS60, GG04}.

The lumped time evolution operator $\bi{H}$ is as before given as the
sum of reproduction and mutation matrix, $\bi{H}=\bi{R+M}$. In the
lumped system, the evolution equation takes the same form as before
(see equation (\ref{1})).

\subsection{Reversibility of $\bi{M}$}
The equilibrium distribution (or PF-eigenvector) $\bi{\pi}$ of
$\bi{M}$ is given by the number of sequences lumped into each
mutational distance,
\begin{equation}
\pi_\bi{d}={N \choose d_0,d_1,d_2,d_3}=\frac{N!}{d_0!d_1!d_2!d_3!} \;.
\end{equation}
Thus $\pi_{\bi{d+e}_k}=\pi_\bi{d}\, d_0/(d_k+1)$ and similarly for
other mutational steps. Using this, it can easily be verified that
$\bi{M}$ is reversible (i.\ e., the detailed balance holds),
\begin{equation}
\label{reversibility}
\bi{M_{d,d'}\pi_{d'}=M_{d',d}\pi_d} \;,
\end{equation}
and as $\bi{R}$ is diagonal, this holds likewise for $\bi{H}$:
$\bi{H_{d,d'}\pi_{d'}=H_{d',d}\pi_d}$.

\section{Maximum principle}

One of the quantities of main interest is the population mean fitness
in equilibrium, $\bar{r}$, which is given by the leading eigenvalue of
$\bi{H}$. In principle, this can be obtained by Rayleigh's variational
principle. Here, this is however not useful, as the optimization would
have to be carried out over a space of dimension
$(N+1)(N+2)(N+3)/6$. Using the assumption of a permutation-invariant
fitness and the lumped Markov chain, Rayleigh's principle can be
reduced to an optimization over a single scalar in the case of binary
sequences \cite{HRWB02}, or over the three components of the
mutational distance our case of four-state sequences \cite{GG04}. A
general reduction of Rayleigh's principle is derived in
\cite{BBBK}. Here, we shall briefly repeat the derivation of the
maximum principle.

\subsection{Symmetrization of $\bi{H}$}

Due to the reversibility of $\bi{M}$, the time-evolution operator
$\bi{H}$ can be symmetrized by the means of a diagonal transformation,
\begin{equation}
\widetilde{\bi{H}}=\bi{\Pi}^{-1/2}\bi{H\Pi}^{1/2}
\end{equation}
using the mutational equilibrium $\bi{\Pi}={\rm
diag}\{\pi_{\bi{d}}\}$. $\widetilde{\bi{H}}$ has the same spectrum as
$\bi{H}$.

The off-diagonal entries of $\widetilde{\bi{H}}$ are given by the
symmetrized mutation rates $\widetilde{\bi{H}}_{\bi{d,d'}}
=\widetilde{\bi{M}}_{\bi{d,d'}}
=\bi{M_{d,d'}}\sqrt{\frac{\pi_\bi{d'}}{\pi_{\bi{d}}}}
=\sqrt{\bi{M_{d,d'}M_{d',d}}}$.  The symmetrized mutation rates are
thus explicitly given by
\begin{equation}
\label{symmutrat}
\eqalign{ \bi{d}\rightarrow\bi{d}+\bi{e}_k: &
\widetilde{\bi{M}}_{\bi{d}+\bi{e}_k,\bi{d}}
=\frac{\mu_k}{N} \sqrt{d_0(d_k+1)} \\
\bi{d}\rightarrow\bi{d}-\bi{e}_k: &
\widetilde{\bi{M}}_{\bi{d}-\bi{e}_k,\bi{d}}
=\frac{\mu_k}{N} \sqrt{d_k(d_0+1)} \\
\bi{d}\rightarrow\bi{d}+\bi{e}_k-\bi{e}_\ell:\quad &
\widetilde{\bi{M}}_{\bi{d}+\bi{e}_k-\bi{e}_\ell,\bi{d}}
=\frac{\mu_m}{N} \sqrt{d_\ell(d_k+1)} \;.
}
\end{equation}
The diagonal entries of $\bi{H}$ are unchanged by this transformation.

\subsection{Splitting up $\widetilde{\bi{H}}$}

The next step is to split up the symmetrized time-evolution operator
$\widetilde{\bi{H}}$ into the sum of a Markov generator and a diagonal
matrix that contains the remainder,
\begin{equation}
\widetilde{\bi{H}}=\bi{E+F}
\end{equation}
with $\bi{F}$ being the (symmetric) Markov generator and $\bi{E}$
being the (diagonal) remainder.  Clearly, the off-diagonal entries of
$\bi{F}$ are given by the symmetrized mutation rates,
$\bi{F_{d',d}}=\widetilde{\bi{M}}_{\bi{d',d}}$. For $\bi{F}$ to be a
Markov generator, we need $\sum_\bi{d'}\bi{F_{d',d}}=0$, and hence the
diagonal entries are given by
\begin{equation}
\eqalign{
\bi{F_{d,d}}=-\sum_{\bi{d'\neq d}}\bi{F_{d',d}}
=&-\sum_k \mu_k/N \left[\sqrt{d_0(d_k+1)}+\sqrt{d_k(d_0+1)}\right] \\
&-\sum_{k,\ell \atop k>\ell} \mu_m /N 
\left[\sqrt{d_\ell(d_k+1)}+\sqrt{d_k(d_\ell+1)}\right] \;,
}
\end{equation}
using the mutation rates (\ref{symmutrat}). The entries of $\bi{E}$
are given by
\begin{equation}
\bi{E_d}=\bi{R_d}+\bi{M_{d,d}}-\bi{F_{d,d}} \;.
\end{equation}

\subsection{The maximum principle}
In the limit $N\rightarrow\infty$ the matrices $\bi{E}$ and $\bi{F}$
can be approximated by functions $E$ and $F_\xi$
\begin{equation}
\bi{E_d}=E(\bi{x_d})+\Or\left(\frac{1}{N}\right) 
\quad\mbox{and}\quad
\bi{F_{d',d}}=F_{\pm\xi}(\bi{x_d})+\Or\left(\frac{1}{N}\right)\;,
\end{equation}
where the $\bi{x_d}$ are normalized mutational distances
$\bi{x_d}=\bi{d}/N$ with $0\le x_k$, $\sum_{k=1}^3x_k\le 1$ and the
$\xi$ labels the possible directions of mutation $k$ or $+k,-\ell$
with $k,\ell \in \{1,2,3\}$ and $k>\ell$. $F_{\bi{d},\bi{d}'}=0$ for
non-neighbouring sequences $\bi{d},\bi{d}'$.

Using the normalized mutational distances $\bi{x}$, we get also a
normalized version of the mutational distance space,
$\mathcal{S}\subset\mathbb{R}^3$, where the points become
dense in the limit $N\rightarrow\infty$.

The $F_{\pm\xi}(\bi{x})$ are given approximately by the
symmetrized mutation rates as
\begin{equation}
F_{\pm k}(\bi{x})\approx\mu_k\sqrt{x_0x_k} \;,\quad
F_{\pm k \mp\ell}(\bi{x})\approx\mu_m\sqrt{x_\ell x_k}\;,
\end{equation}
with the fraction of the wildtype sites $x_0=1-\sum_kx_k$ and
$\{k,\ell,m\}=\{1,2,3\}$. For $E(\bi{x})$ we have
\begin{equation}
E(\bi{x})=r(\bi{x})-g(\bi{x})
\end{equation}
with the fitness function $\bi{R_d}=r(\bi{x_d})+\Or(\case{1}{N})$ and
the mutational loss function
\begin{equation}
g(\bi{x}) :=\sum_k\mu_k\left[1 -2\sqrt{x_0x_k} 
-2\sum_{\ell,m\neq k \atop \ell>m} \sqrt{x_\ell x_m}\right] \;.
\end{equation}

According to \cite{BBBK}, Theorem 1, the population mean fitness in
equilibrium is then given by the maximum principle
\begin{equation}
\label{maximum principle}
\bar{r}=\sup_\bi{x}\left[r(\bi{x})-g(\bi{x})\right]
+\Or(\case{1}{N})\;,
\end{equation}
where the supremum is assumed at the ancestral mean mutational
distance $\hat{\bi{x}}$, if $E(\bi{x})$ assumes its maximum in the
interior of $\mathcal{S}$ (which is the generic case).

\subsubsection{Validity.}
The maximum principle is exact in the limit $N\rightarrow\infty$ as
well as in the case of a linear fitness function $r(\bi{x})$ and for
unidirectional mutation rates, which is not the case for our mutation
model. For finite $N$ (and non-linear fitness function), it is correct
up to $\Or\left(\case{1}{N}\right)$ \cite{BBBK}.

\section{Mutation thresholds}
\subsection{The phenomenon}
Analogously to the temperature-driven phase transition in physics, for
certain choices of fitness functions one can observe the phenomenon
that the order of the population breaks down at a critical mutation
rate, and for any mutation rates higher than this, the population is
delocalized in sequence space. In the population genetics context,
this phenomenon is known as mutation-driven {\em error threshold}.

We define the phenomenon of the mutation threshold for systems with
infinite sequence length $N\rightarrow\infty$ as a critical mutation
rate $\bi{\mu}_c=(\mu_{1,c}, \mu_{2,c}, \mu_{3,c})$, at which there
is a kink in the equilibrium population mean fitness $\bar{r}$ and a
discontinuity in the ancestral mean mutational distance
$\hat{\bi{x}}$, cf.\ the definition of the fitness-threshold in
\cite{HRWB02}. As this is a collective phenomenon, for finite sequence
length, the transitions are smoothed out, and the critical mutation
rates may be shifted a bit. Using the maximum principle
(\ref{maximum principle}), however, we can detect the discontinuities
in $\hat{\bi{x}}$, as this is exact in the limit $N\rightarrow\infty$.

\begin{figure}
\begin{center}
\includegraphics[width=\textwidth]{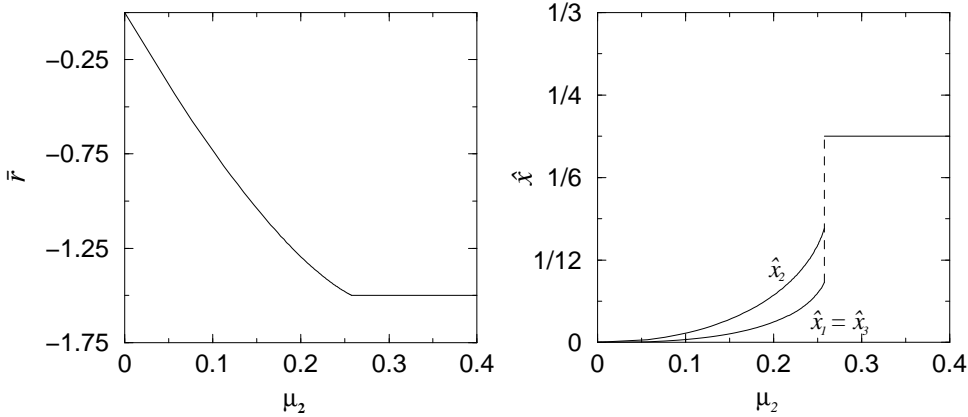}
\end{center}
\caption{\label{Fig3}Example of a phase transition for Kimura 2
parameter mutation scheme with $\mu=2\mu_2$ and a quadratic symmetric
fitness $r(\bi{x})=-\sum_kx_k+\sum_{k\le\ell}x_kx_\ell$.}
\end{figure}

Figure \ref{Fig3} shows an example of an error threshold for the
Kimura 2 parameter mutation model with $\mu=2\mu_2$ and a quadratic
symmetric fitness
\begin{equation}
r(\bi{x})=-\sum_{k=1}^3x_k+\sum_{k,\ell=1 \atop k\le\ell}^3 
x_k x_\ell\;.
\end{equation}
As the fitness function is symmetric with respect to the $x_k$, and
the K2P mutation scheme breaks the symmetry only between $x_2$ 
vs.\ $x_1,x_3$, in equilibrium we have $\hat{x}_1=\hat{x}_3$.  

In Figure \ref{Fig3}, one can see the decline in population mean
fitness for mutation rates below the critical rate, which goes in line
with an increase in ancestral mean mutational distance
$\hat{\bi{x}}$. At the critical mutation rate, $\mu_{2,c}\approx0.26$,
the population mean fitness has a kink, and the ancestral mean
mutational distance $\hat{\bi{x}}$ jumps to the completely random
mutational distance with
$\hat{x}_0=\hat{x}_1=\hat{x}_2=\hat{x}_3=\case{1}{4}$, where it stays
for all higher mutation rates, and consequently, the population mean
fitness $\bar{r}$ stays constantly at its minimum value.

\subsection{Choice of fitness functions}
Error thresholds do not occur for all fitness functions, but only for
some classes. For two-state models, there exist criteria for fitness
functions to give rise to error thresholds \cite{HRWB02}. For our
four-state model, it is plausible to assume that the situation is
similar.

It is easy to verify for the four-state model, that there can be no
error thresholds for linear fitness functions.  For a fitness function
to give rise to error thresholds, one needs at least the complexity of
a quadratic function. In the following, we will investigate two
examples of fitness functions displaying error thresholds, namely a
quadratic fitness function and a truncation selection, where all
genotypes with less than a critical number of mutations are equally
fit and all others equally unfit.

For simplicity, we will limit ourselves in our examples to the Kimura
2 parameter (K2P) and Jukes-Cantor (JC) mutation schemes as
simplifications of the full Kimura 3ST mutation scheme, and restrict
the choice of fitness functions to those that are symmetric with
respect to the $x_k$, $k\in\{1,2,3\}$.

\subsection{Quadratic symmetric fitness function}
It can be shown that for quadratic fitness functions with positive
epistasis, i.\ e., fitness functions with negative second derivatives,
there exist no phase transitions (cf.\ \cite{HWB01}). Looking only at
quadratic symmetric fitness functions with negative epistasis (or
positive second derivative), a fairly general form is
\begin{equation}
\label{fitfunc}
r(\bi{x})=c\sum_{k=1}^3x_k
+\sum_{k,\ell=1 \atop k\le\ell}^3 x_k x_\ell \;.
\end{equation}
Here, the parameter $c$ is used to tune the linear part relative to
the quadratic term. The only possible generalization within our
restricted setup would be to give different coefficients for the pure
quadratic terms $x_k^2$ and the mixed quadratic terms $x_k x_\ell$
with $k\neq\ell$. We will however concentrate on the fitness given in
equation (\ref{fitfunc}).

In general, this fitness is symmetric with respect of the $x_k$,
$k\in\{1,2,3\}$. For $c=-1$, the symmetry of the fitness function is
even higher and includes the fraction of the wildtype-sites,
$x_0=1-\sum_{k=1}^3x_k$. This is the fitness function that has been
used for the example of an error threshold in Figure \ref{Fig3}.

As the K2P mutation model has an inherent symmetry between $x_1$ and
$x_3$, and between $x_2$ and $x_0$, respectively, and the fitness
function (\ref{fitfunc}) for arbitrary $c$ is symmetric with respect
to the $x_k$, $k\in\{1,2,3\}$, it is evident that the equilibrium
solution for the ancestral mean mutational distance mirrors that
symmetry with $\hat{x}_1=\hat{x}_3$ (and in the case of $c=-1$, also
$\hat{x}_2=\hat{x}_0$).
\begin{figure}
\begin{center}
\includegraphics[width=0.7\textwidth]{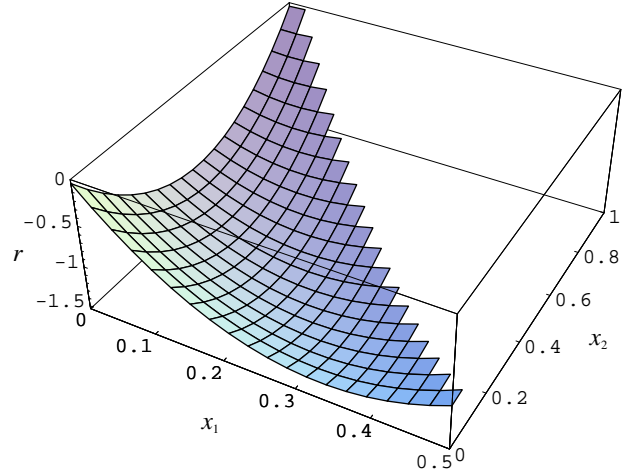}
\end{center}
\caption{\label{Fig4}The quadratic symmetric fitness function
(\ref{fitfunc}) for $c=-1$ as a projection onto the relevant subspace
with $x_1=x_3$.}
\end{figure}
Figure \ref{Fig4} shows the quadratic symmetric fitness function
(\ref{fitfunc}) for $c=-1$ as a projection onto the relevant subspace
where $x_1=x_3$. The additional symmetry in the case $c=-1$ can be
seen from the fact that there are two equally high maxima at
$(x_1,x_2)=(0,0)$ and $(x_1,x_2)=(0,1)$.

\subsubsection{Phase diagrams.}

We have examined the phase diagrams for the K2P mutation model with
the quadratic symmetric fitness function (\ref{fitfunc}) for the
possible combinations of mutation rates and the parameter $c$.

\begin{figure}
\begin{center}
\includegraphics[width=0.8\textwidth]{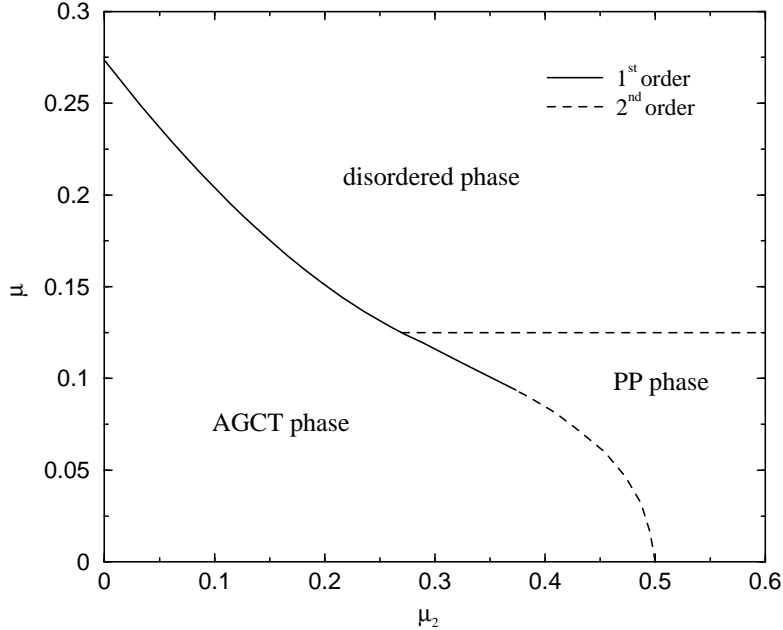}
\end{center}
\caption{\label{Fig5}Phase diagram for the K2P mutation model and the
quadratic symmetric fitness function (\ref{fitfunc}) with $c=-1$. At
the first order phase transition (solid line) the ancestral mean
mutational distance $\hat{\bi{x}}$ jumps and the population mean
fitness $\bar{r}$ has a kink, whereas at the second order phase
transitions (dashed lines) this kink in $\bar{r}$ is reduced to a kink
in its first derivative and the jump in $\hat{\bi{x}}$ is reduced to
an infinite derivative at the critical mutation rate. }
\end{figure}
For the case $c=-1$, the phase diagram is shown in Figure \ref{Fig5},
cf.\ \cite{HWB01}, where a different notation, but the same fitness
and mutation model are used. Here, we can identify 3 different phases:
\begin{itemize}
\item The AGCT phase. The population is essentially ordered, i.\ e.,
the population distribution is localized in sequence space. 
\item The disordered phase. The population is completely random, the
population distribution is the equidistribution in sequence space, the
ancestral mean mutational distance is given by
$\hat{x}_0=\hat{x}_1=\hat{x}_2=\hat{x}_3=\case{1}{4}$.
\item The PP phase. A partially ordered phase, which only
differentiates between purines and pyrimidines.  In the ancestral
distribution, there are two peaks that are equidistributions with
respect to $x_1,x_3$-- and $x_0,x_2$--direction, respectively, but
localized in the other directions. This phase only exists in the case
$c=-1$.
\end{itemize}

\begin{figure}
\begin{center}
\includegraphics[width=0.8\textwidth]{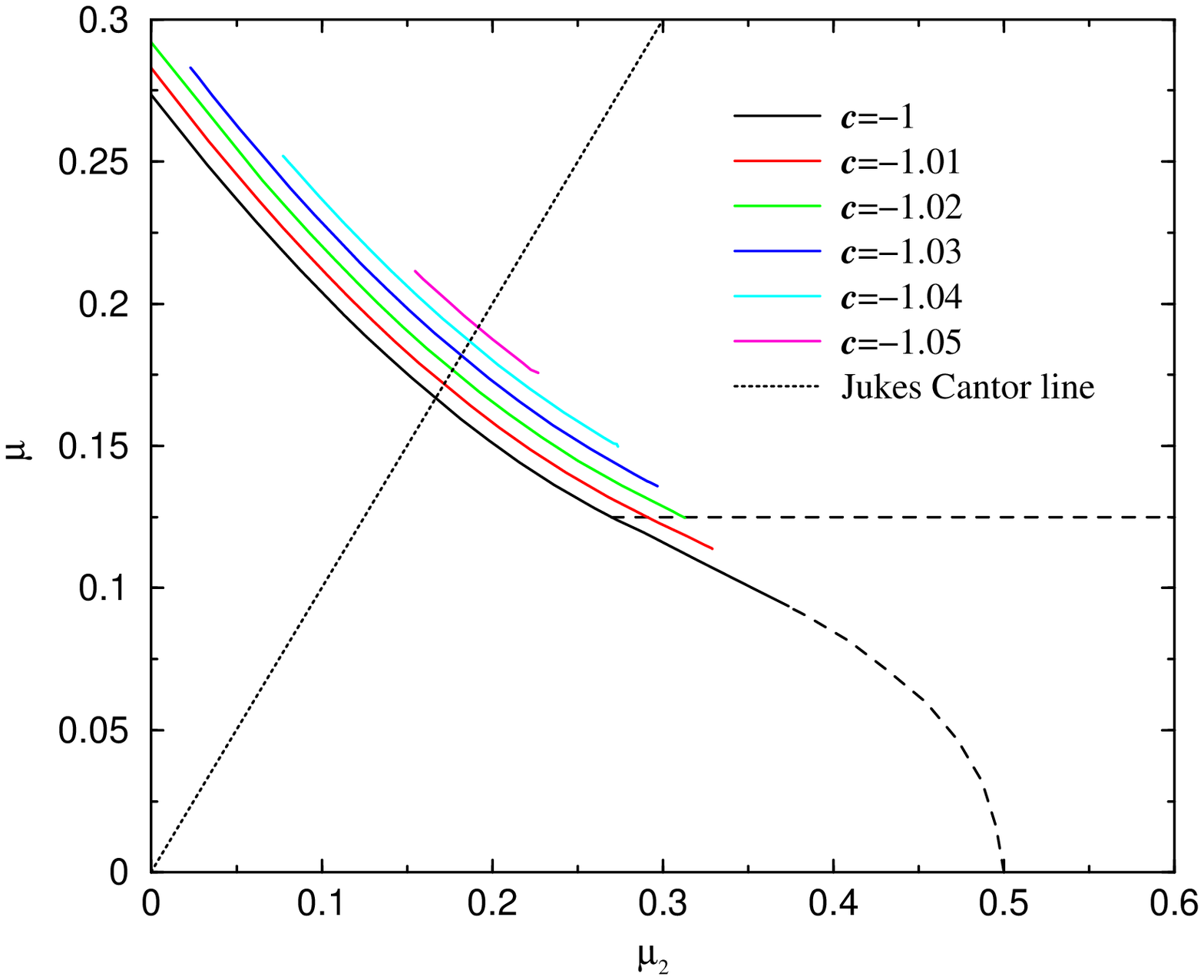}
\end{center}
\caption{\label{Fig6}Phase diagram for the K2P mutation model and the
quadratic symmetric fitness function (\ref{fitfunc}) with $c\le-1$. }
\end{figure}

\begin{figure}
\begin{center}
\includegraphics[width=0.8\textwidth]{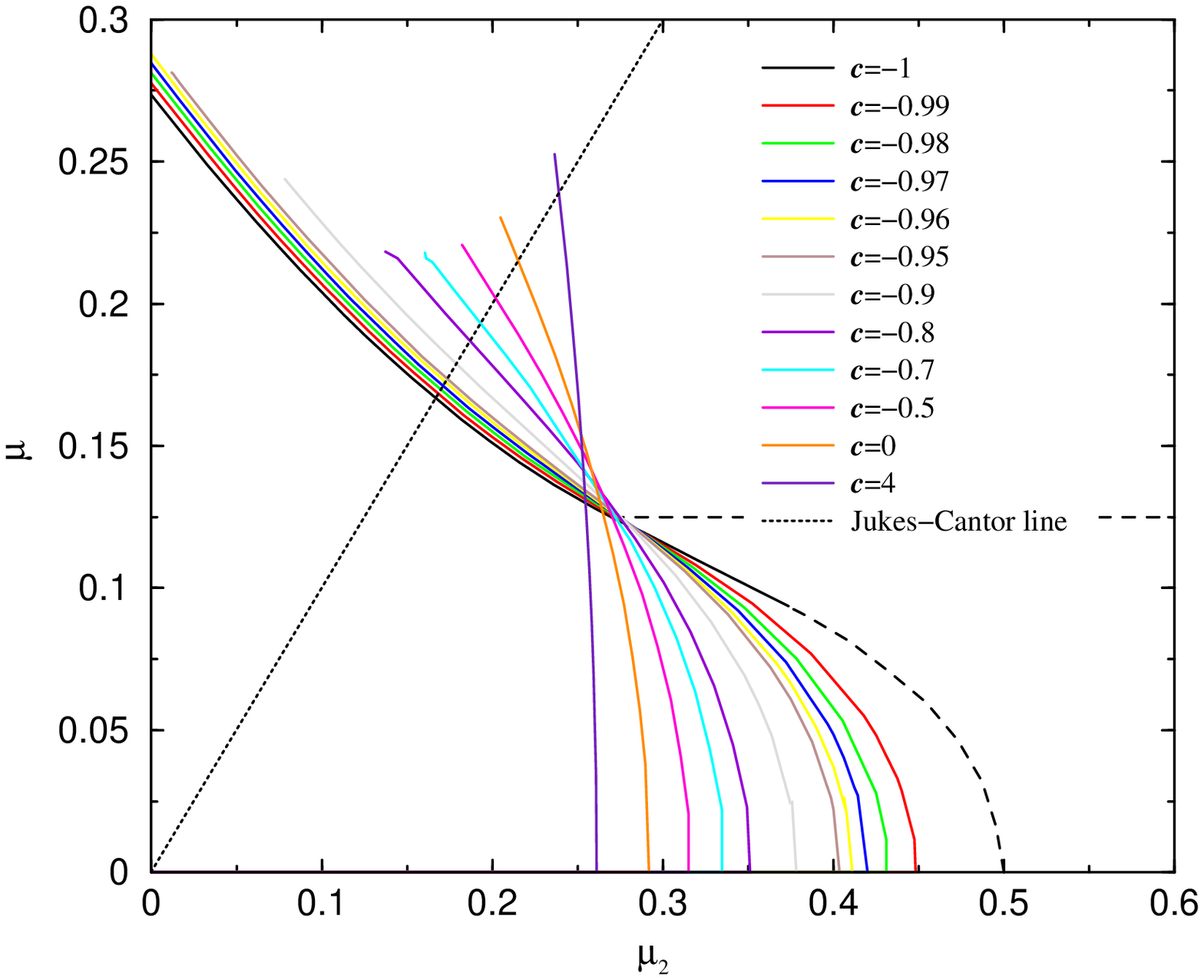}
\end{center}
\caption{\label{Fig7}Phase diagram for the K2P mutation model and the
quadratic symmetric fitness function (\ref{fitfunc}) with $c>-1$. }
\end{figure}

If the symmetry between $x_0$ and $x_2$ is broken, i.\ e., $c\neq -1$,
the PP phase disappears, see Figures \ref{Fig6} and \ref{Fig7} for
$c\le -1$ and $c\ge -1$, respectively.

For $c<-1$, cf.\ Figure \ref{Fig6}, the second order phase transitions
disappear immediately with the broken symmetry, the 1st order phase
transition lines shrink with decreasing $c$ while shifting to slightly
higher mutation rates and disappear for any $c\le -1.06$. The
disappearance of the 2nd order transitions implies a disappearance of
the disordered and the PP phase, leaving only an AGCT phase with
varying degrees of order.

In the case $c>-1$, cf.\ Figure \ref{Fig7}, the PP phase merges with
the disordered phase, such that only the AGCT and disordered phases
remain. The second order phase transition between AGCT and PP phase in
the case $c=-1$ is transformed to a first order transition.

\subsubsection{Finite size effects.}
Using the maximum principle (equation (\ref{maximum principle})), only
the population mean fitness $\bar{r}$ and the ancestral mutational
distance $\hat{\bi{x}}$ are accessible, which is sufficient to detect
phase transitions. For small sequence length, it is however feasible
to calculate $\bar{r}$ as largest eigenvalue and the population and
ancestral distributions $\bi{p}$ and $\bi{a}$ through the
corresponding eigenvector of $\widetilde{\bi{H}}$.

\begin{figure}
\begin{center}
\includegraphics[width=\textwidth]{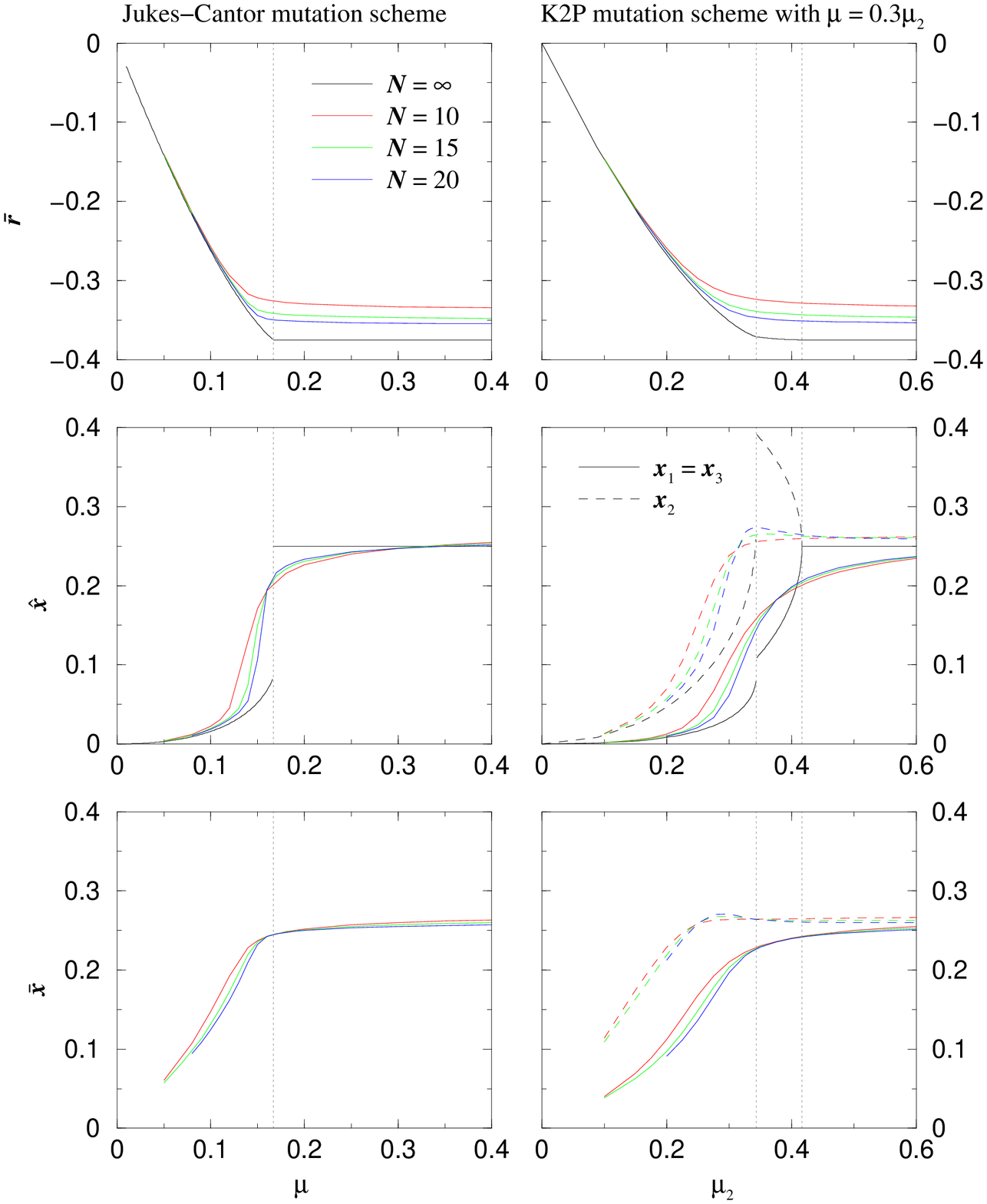}
\end{center}
\caption{\label{Fig8}Comparison of population mean fitness $\bar{r}$,
ancestral mean mutational distance $\hat{\bi{x}}$ and population mean
mutational distance $\bar{\bi{x}}$ for different sequence length and
the infinite system with fitness function (\ref{fitfunc}) with
$c=-1$. Left: JC mutation scheme, right: K2P mutation scheme
with $\mu=0.3\mu_2$. The locations of the phase transitions are marked
as vertical dotted lines. }
\end{figure}
Figure \ref{Fig8} shows results obtained this way for different finite
sequence lengths using the quadratic fitness function (\ref{fitfunc})
with parameter $c=-1$ compared with the results obtained via maximum
principle (\ref{maximum principle}), which are exact for an infinite
system. On the left, the JC mutation model was used, and
consequently, the mean mutational distances in all three directions
coincide. On the right, the K2P mutation model with $\mu=0.3\mu_2$ was
used. Here, the mean mutational distances in $x_1$ and $x_3$ direction
coincide, but differ from those in $x_2$ direction.

In Figure \ref{Fig8} one can clearly see how the phase transitions,
which are very sharp in the infinite system, are smoothed out for
finite sequence length. Especially the second order phase transition
in the K2P model cannot be detected at the sequence lengths considered
here.

\subsubsection{Distributions.}

\begin{figure}
\begin{center}
\includegraphics[width=\textwidth]{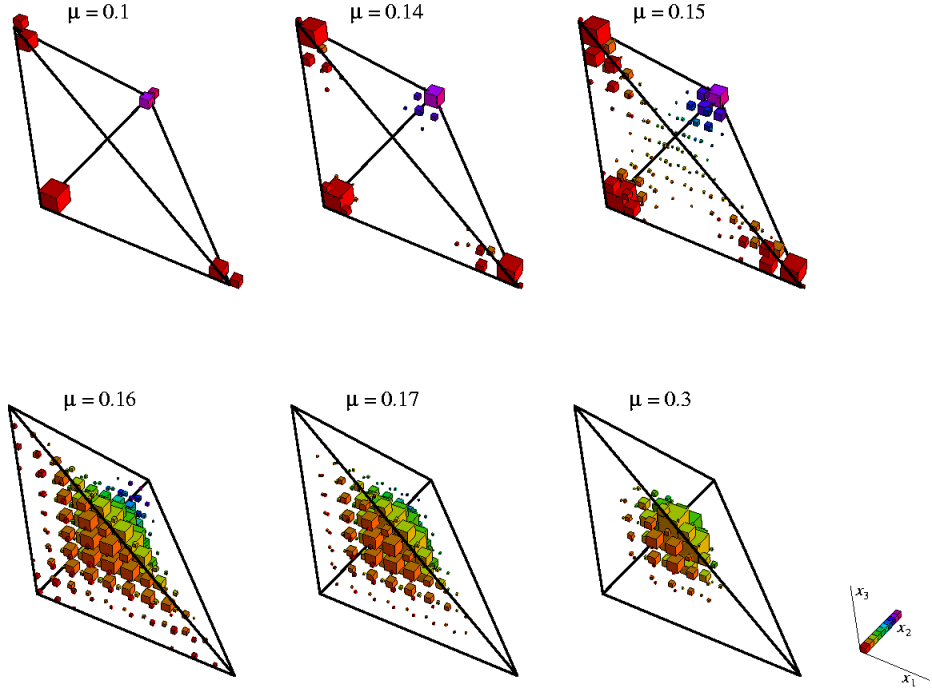}
\end{center}
\caption{\label{Fig9}Ancestral distribution $\bi{a}$ for the JC
mutation model with quadratic symmetric fitness (\ref{fitfunc}) and
$c=-1$ for selected mutation rates for a sequence length of
$N=20$. The frequency of each type $\bi{d}$ is symbolized by a cube of
proportional size. Colours indicate the position along the $x_2$
direction. For easier recognition, each cube corresponds to a block of
8 data points. }
\end{figure}

\begin{figure}
\begin{center}
\includegraphics[width=\textwidth]{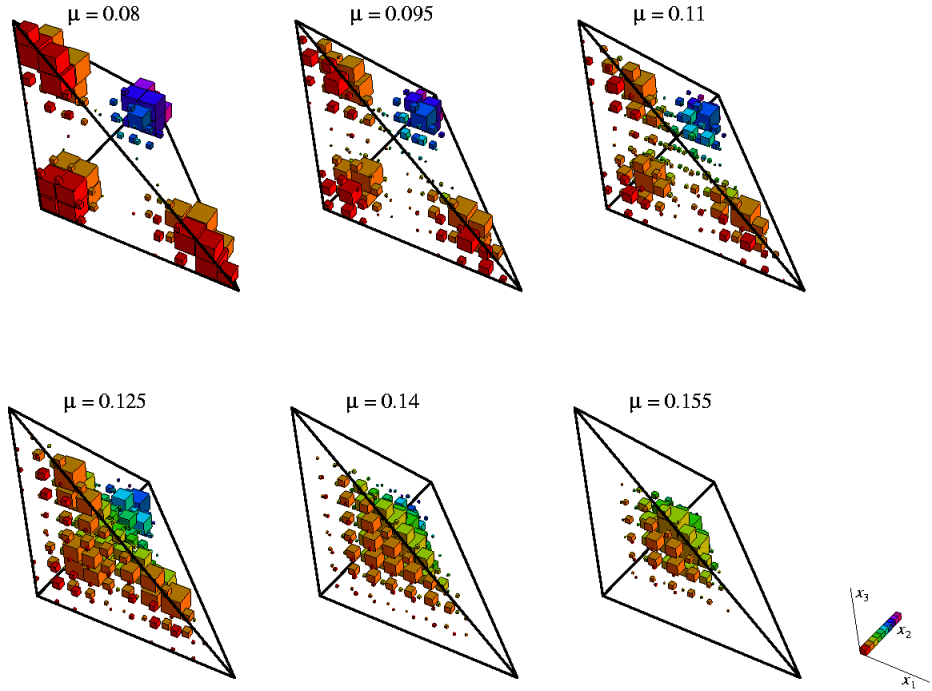}
\end{center}
\caption{\label{Fig10}Population distribution $\bi{p}$ for the JC
mutation model with fitness (\ref{fitfunc}) and $c=-1$ for selected
mutation rates for a sequence length of $N=20$. For easier
recognition, each cube corresponds to a block of 8 data points.}
\end{figure}

\begin{figure}
\begin{center}
\includegraphics[width=\textwidth]{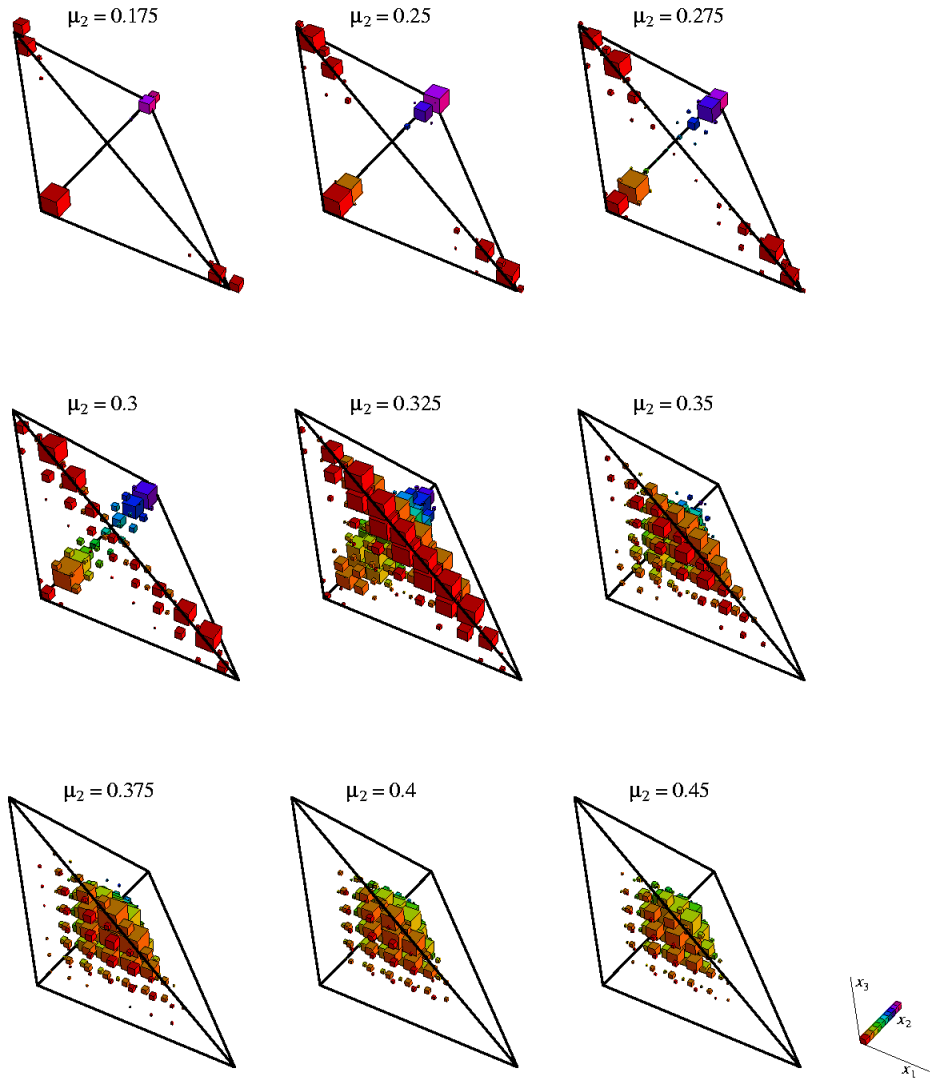}
\end{center}
\caption{\label{Fig11}Ancestral distribution $\bi{a}$ for the K2P
mutation model with $\mu=0.3 \mu_2$ with fitness (\ref{fitfunc}) and
$c=-1$ for selected mutation rates for a sequence length of
$N=20$. For easier recognition, each cube corresponds to a block of 8
data points.}
\end{figure}

The results in Figure \ref{Fig8} have been obtained by calculating the
population and ancestral distributions explicitly. Examples of these
are visualized in Figures \ref{Fig9} to \ref{Fig11}.  Although for the
mean mutational distance we have $x_1=x_3$ and therefore can reduce
the sequence space to the 2-dimensional subspace, sequences with
$x_1\neq x_3$ do occur in the equilibrium distributions with non-zero
frequency, and thus we need the full three-dimensional mutational
distance space $\mathcal{S}$ for the visualization. In Figures
\ref{Fig9} to \ref{Fig11}, the frequency of each point $\bi{d}$ is
visualized as a cube of proportional size. For easier recognition,
each cube corresponds to a block of 8 data points. The colours of the
cubes indicate their position along the $x_2$ direction.

Figure \ref{Fig9} shows the ancestral distribution of the JC mutation
model with the quadratic symmetric fitness function (\ref{fitfunc})
with parameter $c=-1$. The phase transition, which happens around a
mutation rate of $\mu\approx 0.16$, is clearly visible. For lower
mutation rates, the population is in the AGCT phase and the ancestral
distribution is localized. For mutation rates beyond the threshold,
the population is in the disordered phase and the ancestral
distribution is the equidistribution in sequence space. Note that the
term equidistribution refers to the $4^N$-dimensional sequence space
$\mathfrak{S}$, and thus we have a multinomial distribution in the
depicted mutational distance space $\mathcal{S}$.

In Figure \ref{Fig10}, the population distributions for the same model
and parameter values as in Figure \ref{Fig9} are shown. The population
distribution goes through the same stages as the ancestral
distribution, but at lower mutation rates, and in contrast to the
ancestral distribution, the transition is smooth, as can be seen
already in Figure \ref{Fig8}.

Figure \ref{Fig11} shows the ancestral distribution for the K2P
mutation model with $\mu=0.3 \mu_2$ and fitness function
(\ref{fitfunc}) with $c=-1$. Here, the population starts in the AGCT
phase (mutation rates $\mu_2 < 0.3$) and goes through the PP phase (at
$\mu_2=0.325$), where the distribution constists of two ``rods'',
which are each an equidistribution with respect to the $x_2,x_0$ or
the $x_3,x_1$ direction, respectively, and localized with respect to
the remaininig directions.  For high mutation rates, we find again the
equidistribution of the disordered phase ($\mu_2>0.4$).

\clearpage

\subsection{Truncation selection}

Another fitness we are interested in is truncation selection, a case
of extreme epistasis, both positive and negative. We use it in the
form
\begin{equation}
\label{truncsel}
r(\bi{x})=\left\{\begin{array}{cl}
1 & \mbox{if } \sum_k x_k\le 3x_c\\[1ex]
0 & \mbox{if } \sum_k x_k>3x_c \quad .
\end{array}\right.
\end{equation}
\begin{figure}
\begin{center}
\includegraphics[width=0.4\textwidth]{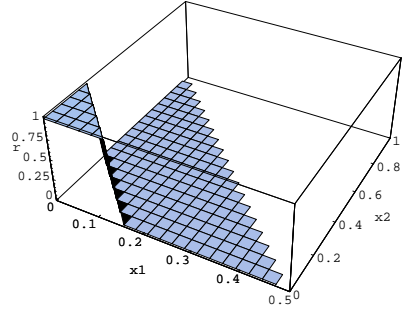}
\end{center}
\caption{\label{Fig12}Truncation selection (\ref{truncsel}) for
$x_c=0.1$ as a projection on the subspace $x_1=x_3$. }
\end{figure}
Figure \ref{Fig12} shows a projection of the truncation selelction
(\ref{truncsel}) with $x_c=0.1$ onto the subspace $x_1=x_3$. 

This is a generalization of the single-peaked landscape, where one
single genotype is of high fitness and all others are equally
unfit. The single peaked landscape has been widely used as an
oversimplistic toy model. It has first been suggested in the context
of prebiotic evolution \cite{Eig71} and corresponds to $x_c=0$ in our
setting. Truncation selection with non-zero $x_c$, however, is a
standard model in biology, e.\ g.\ \cite{Kon88}.

\subsubsection{Phase diagram.}
\begin{figure}
\centerline{
\includegraphics[width=0.5\textwidth]{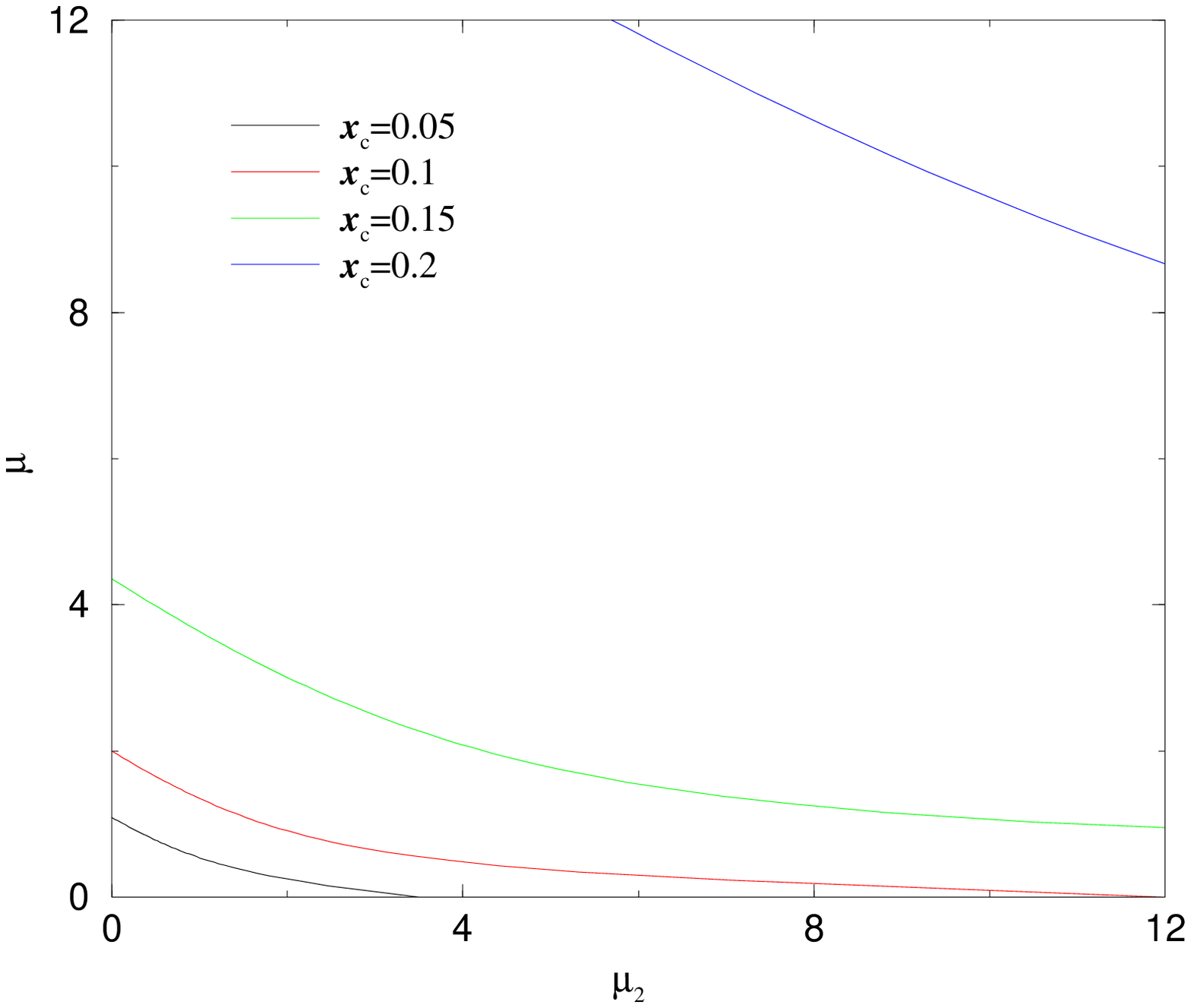}
\includegraphics[width=0.5\textwidth]{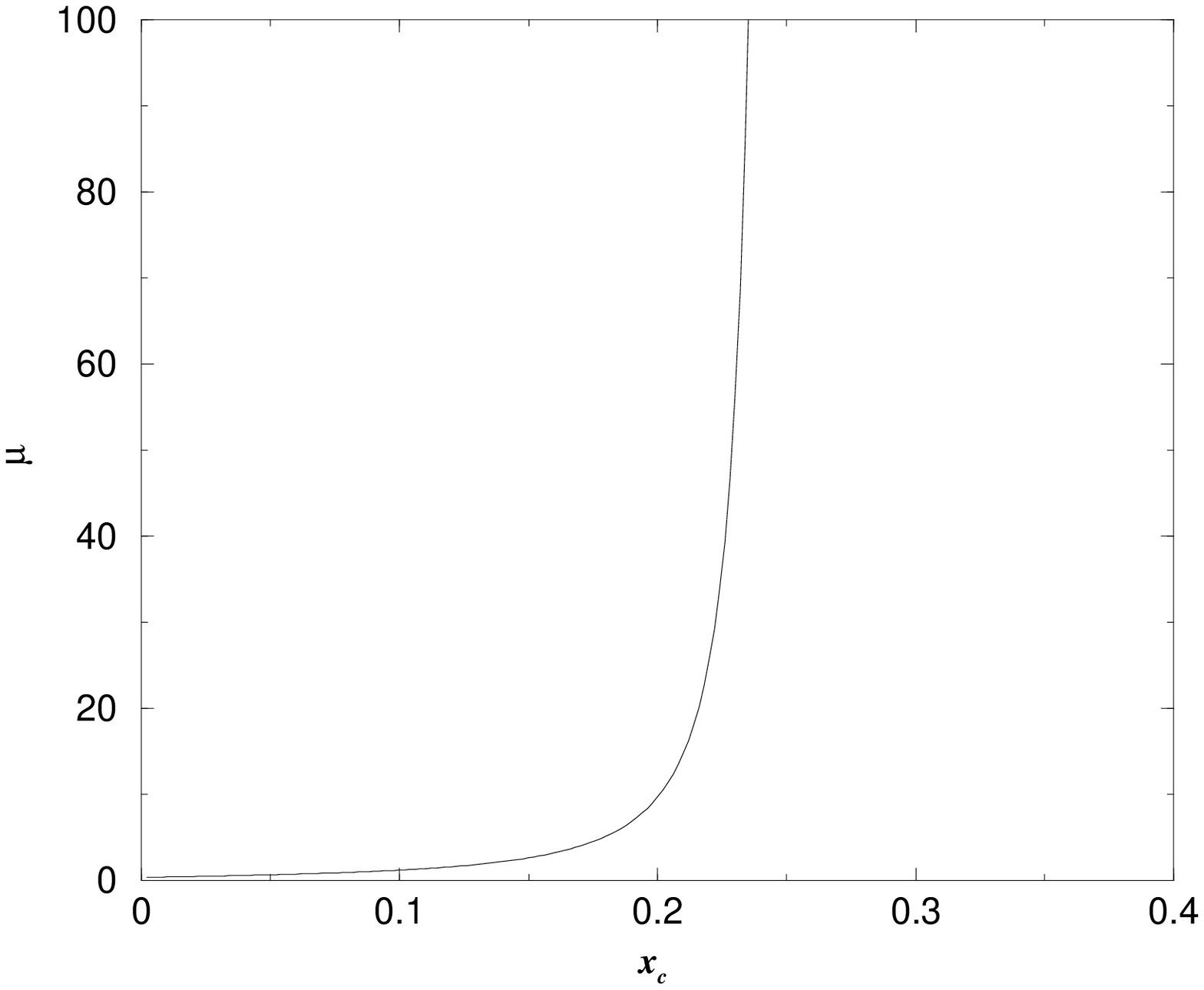}}
\caption{\label{Fig13}Left: Phase diagram for truncation selection
(\ref{truncsel}) with different values of $x_c$. 
Right: Critical mutation rate $\mu_c$ against $x_c$ for truncation
selection (\ref{truncsel}) and the JC mutation model.}
\end{figure}
On the left, Figure \ref{Fig13} shows the phase diagram for truncation
selection (\ref{truncsel}) with different values of $x_c$. Here, we
find error thresholds for all values of $x_c<\case{1}{4}$. For values
of $x_c\ge \case{1}{4}$, the point of the mutation equilibrium,
$\bi{x}=(\case{1}{4},\case{1}{4},\case{1}{4})$, is included in the
high plateau and thus the population will be in mutation equilibrium
and at optimal fitness simultaneously for any mutation rate. On the
right, Figure \ref{Fig13} shows the dependence of the critical
mutation rate on $x_c$ in the JC mutation model. At $x_c=\case{1}{4}$,
the critical mutation rate diverges.

\subsubsection{Finite size effects.}
\begin{figure}
\begin{center}
\includegraphics[width=\textwidth]{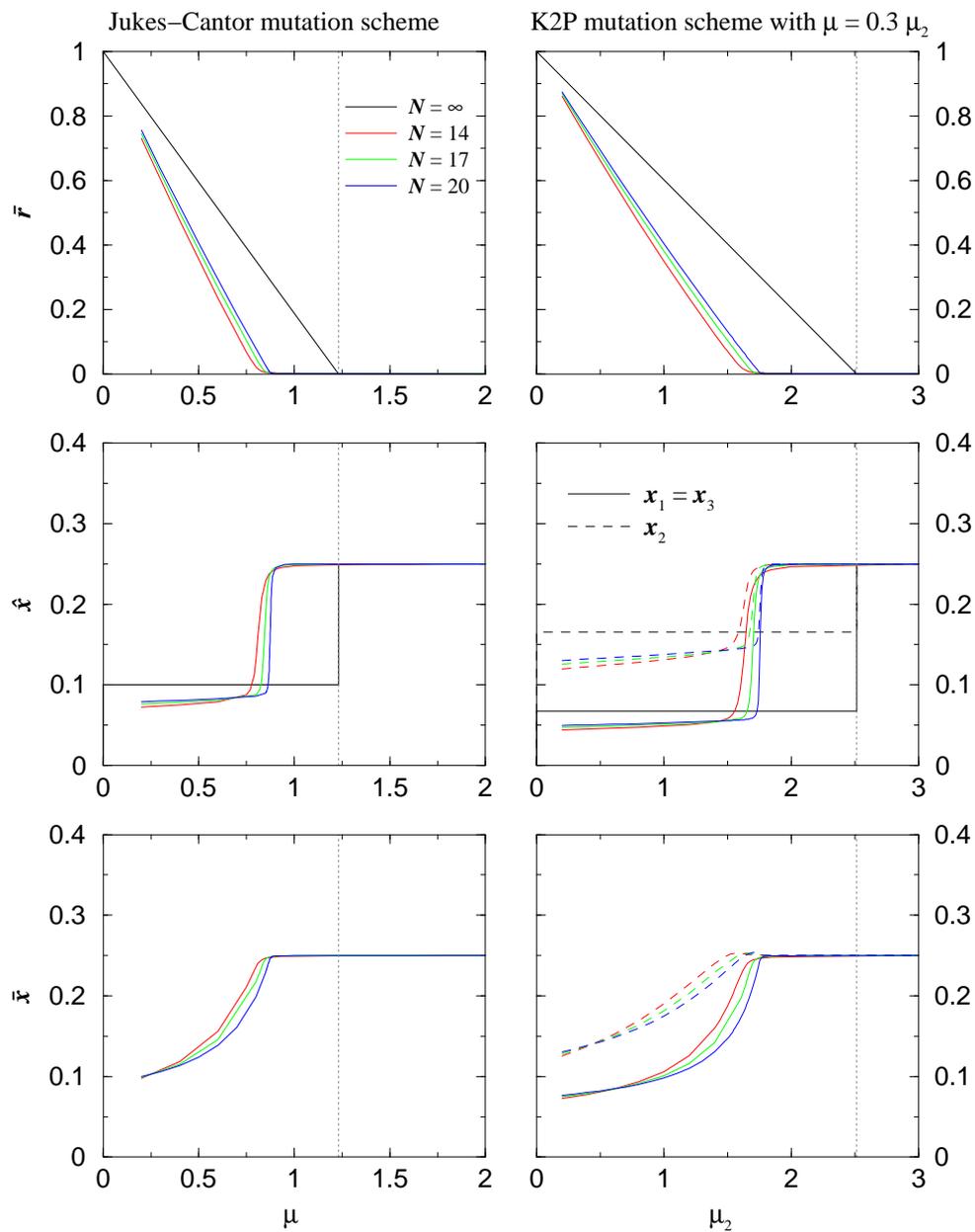}
\end{center}
\caption{\label{Fig14}Comparison of population mean fitness $\bar{r}$,
ancestral mean mutational distance $\hat{\bi{x}}$ and population mean
mutational distance $\bar{\bi{x}}$ for different sequence lengths and
the infinite system with truncation selection (\ref{truncsel}) with
$x_c=0.1$. Left: JC mutation scheme, right: K2P mutation
scheme with $\mu=0.3\mu_2$. The locations of the error thresholds in
the infinite system are marked as vertical dotted lines. }
\end{figure}
Figure \ref{Fig14} compares the results for truncation selection
obtained via the maximum principle (\ref{maximum principle})
($N=\infty$) with those obtained by calculating the largest eigenvalue
and corresponding eigenvector of the symmetrized time-evolution
operator $\widetilde{\bi{H}}$ for finite sequence length analogously
to the data shown in Figure \ref{Fig8} for the quadratic symmetric
fitness (\ref{fitfunc}). On the left, one sees the data for the JC
mutation scheme, whereas on the right, the results for the K2P model
with $\mu=0.3\mu_2$ are shown. Here, it is obvious that even for
rather small sequence length, the error thresholds are very
sharp. However, keeping in mind the discontinuity of the truncation
selection, this is not too surprising.  One can see that the location
of the error threshold does depend on the sequence length: The smaller
the sequence length $N$, the more the error threshold is shifted to
lower mutation rates. So although the sequence lengths considered here
are large enough to warrant a sharp error threshold, they are too
small to predict the location of the error threshold in the infinite
system.

\subsubsection{Distributions.}

\begin{figure}
\begin{center}
\includegraphics[width=\textwidth]{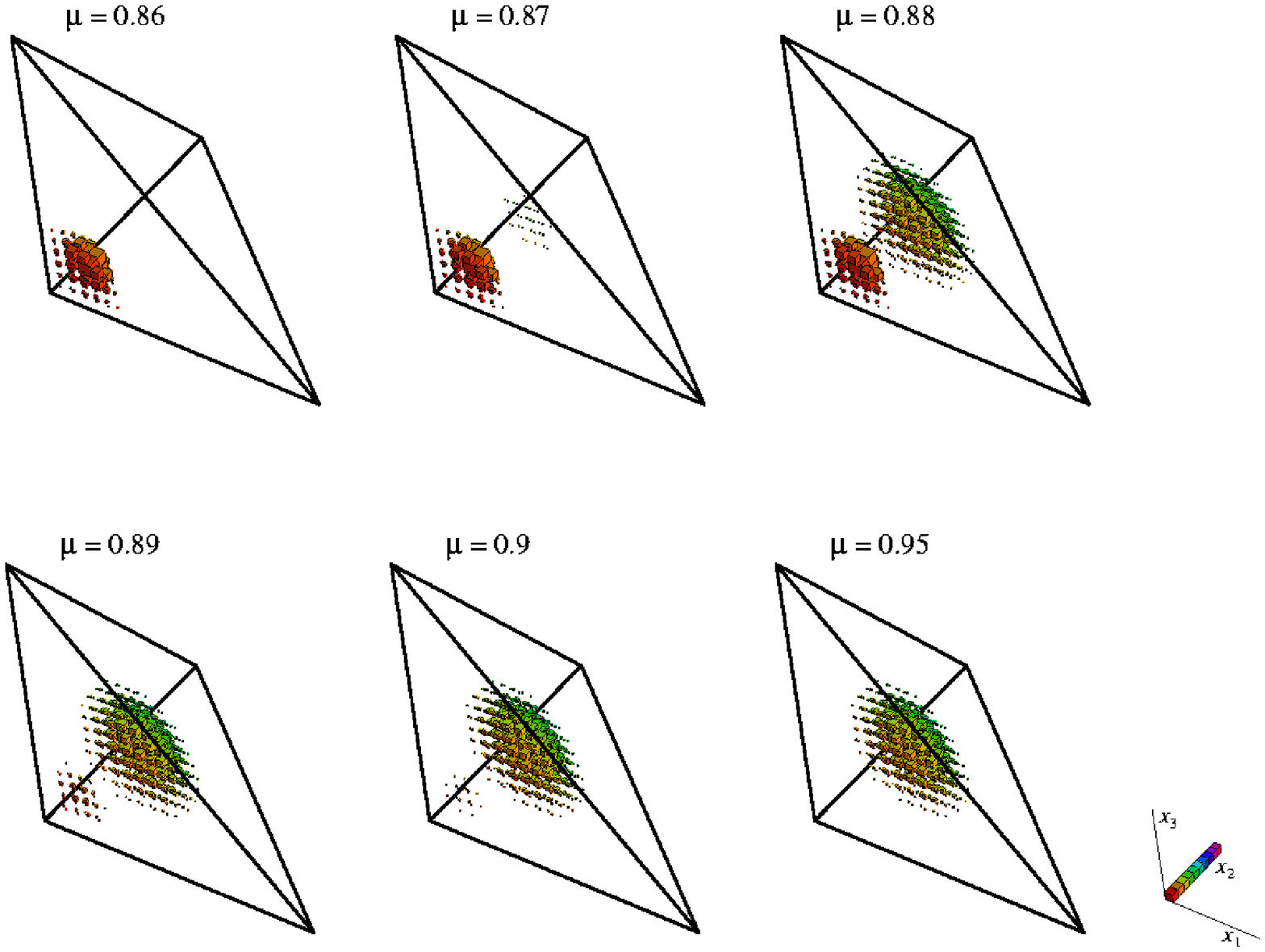}
\end{center}
\caption{\label{Fig15}Ancestral distribution $\bi{a}$ for the JC
mutation model with truncation selection (\ref{truncsel}) and
$x_c=0.1$ with a sequence length of $N=20$ for mutation rates $\mu$
close to the error threshold.}
\end{figure}

\begin{figure}
\begin{center}
\includegraphics[width=\textwidth]{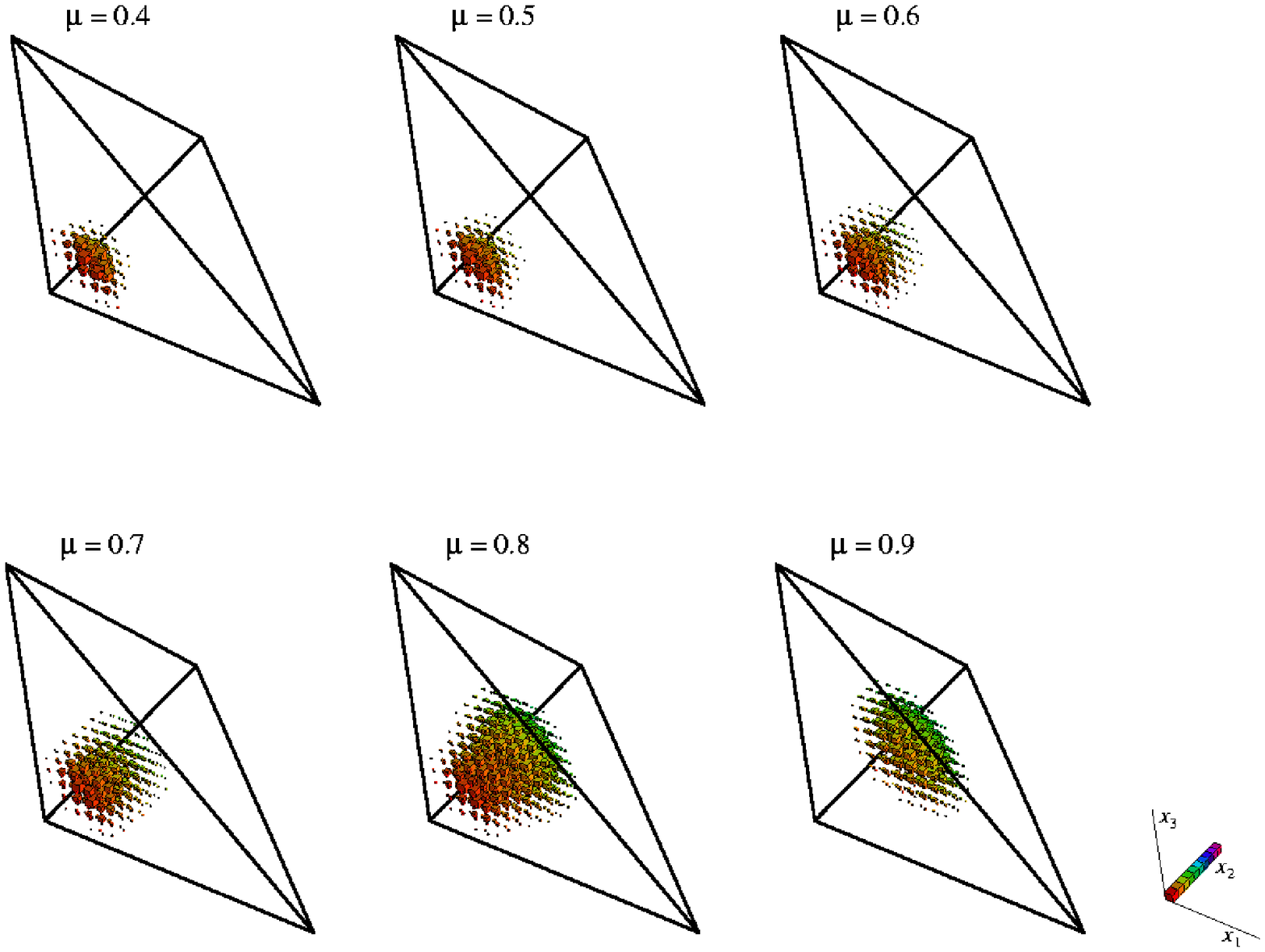}
\end{center}
\caption{\label{Fig16}Population distribution $\bi{p}$ for the JC
mutation model with truncation selection (\ref{truncsel}) and
$x_c=0.1$ with a sequence length of $N=20$ for varying mutation rates
$\mu$.}
\end{figure}

In the same way as for the quadratic symmetric fitness function
(\ref{fitfunc}), the ancestral and population distributions have been
calculated for the truncation selection (\ref{truncsel}) and are shown
in Figures \ref{Fig15} and \ref{Fig16}. In contrast to Figures
\ref{Fig9} to \ref{Fig11}, which are the equivalent diagrams for the
quadratic symmetric fitness function, every point in the sequence
space is displayed as a separate cube, whose size is proportional to
the fraction of individuals having that type in the ancestral and
population distributions, respectively.

Figure \ref{Fig15} shows the ancestral distribution of a system with
sequence length $N=20$ and the JC mutation scheme for
mutation rates close to the error threshold. For mutation rates below
the threshold ($\mu\le 0.86$), the ancestral distribution is an
equidistribution in the ``fit'' part of the sequence space, where
$\sum_k x_k\le x_c$ and thus $r=1$. For mutation rates above the
threshold ($\mu\ge 0.95$), the ancestral distribution is the
equidistribution in the whole sequence space. The transition at the
threshold is however interesting. Here, the distribution does not move
smoothly from one equidistribution to the other, but in the
intermediate states, the ancestral distribution is a superposition of
the two equidistributions with an increasing proportion of the
equidistribution on the whole sequence space.

Figure \ref{Fig16} shows the population distribution for a system with
the same parameters as shown in Figure \ref{Fig15}, i.\ e., JC
mutation model and truncation selection (\ref{truncsel}) with
$x_c=0.1$ and sequence length $N=20$. Here, the transition from a
localized distribution for small mutation rates to the
equidistribution in sequence space looks far smoother (even apart from
the far broader range of mutation rates over which it happens). The
distribution as a whole shifts its centre until it reaches the
equidistribution.

As the situation is very similar for general K2P mutation model (apart
from a difference in mutation rates, cf.\ Figure \ref{Fig13}, left),
no distributions for a case with different mutation rates are shown.

\section{Conclusion}

In this paper we presented the model for sequence evolution introduced
in \cite{HWB01}, which takes into account the four-letter structure of
DNA sequences. In \cite{GG04}, a maximum principle to determine the
population mean fitness has been derived for this model, which is
equivalent to the principle of minimum free energy in physical
systems.  Here, we used this maximum principle to investigate the
phenomenon of error thresholds that are driven by mutation rate for a
number of models.

The general mutation model in our setup is the Kimura 3ST mutation
scheme, where it is accounted for one mutation rate for transitions
and two mutation rates for the different types of transversions. In
the analysis of the phase diagrams we focused on 2 simplifications of
this model: Firstly, the Kimura 2 parameter model, where the two
mutation rates for transversions are taken to be identical, and
secondly, the Jukes-Cantor mutation model as a special case of the
Kimura 2 parameter model where all three mutation rates coincide.

Inherent in our setup, there is a restriction to permutation-invariant
fitness functions, where the fitness depends only on the number of
mutations, i.\ e., the mutational distance $\bi{d}$, not on their
position in the sequence. In principle, it is possible to consider any
fitness function from this permutation-invariant class within our
framework. Not all possible fitness functions give rise to error
thresholds, for instance, with non-epistatic fitness functions, i.\
e., functions that are linear in the mutational distance $\bi{d}$,
there exist no error thresholds.  Here, we have focused on two simple
examples of fitness functions that do cause error thresholds, namely
(i) a fairly general quadratic symmetric (with respect to the
components of the mutational distance $d_i$) fitness function and (ii)
a truncation selection.

For these fitness functions, error thresholds can be found for certain
parameter values. It has been argued (e.\ g.\ \cite{Cha90}), that the
error threshold phenomenon, which was first described for a model with
single-peaked landscape, might be an artefact of this highly
unrealistic fitness function. Our results show, however, that they are
not limited to this fitness, but occur for different types of fitness
functions both smooth (the quadratic fitness) and discontinuous
(truncation selection) ones. A certain degree of complexity is however
needed for error thresholds, as there are none for a linear
fitness. Therefore one can expect that more realistic fitness
landscapes, which are more rugged and not of the permutation-invariant
type, might show a rich error threshold behaviour. Hence error
thresholds are a phenomenon that plays a role in evolution, as supported by 
recent experimental data \cite{DH97, CCA01}.

Apart from the ordered and disordered phases, which also occur in
two-state models, a partially ordered phase has been observed
\cite{HWB01}, which differentiates only between purines and
pyrimidines. This phase is directly dependent on the four-state nature
of our model. However, as our analysis has shown, it is as well
dependent on a certain symmetry of the fitness function, which comes
naturally in the equivalent physical systems, but seems rather
artificial in the biological setting, indicating that this phase,
interesting though it may be, is of little significance in biology.

\ack 
It is our pleasure to thank E.\ Baake and M.\ Baake for helpful
discussions. We gratefully acknowledge support by the British Council
under the Academic Research Collaboration (ARC) Programme, Project No
1213.


\section*{References}
\begin{thebibliography}{10}

\bibitem{BBBK}
E.\ Baake, M.\ Baake, A.\ Bovier, and M.\ Klein.
\newblock An asymptotic maximum principle for essentially linear evolution
  models.
\newblock 2004.
\newblock Preprint q-bio.PE/0311020.

\bibitem{BBW97}
E.\ Baake, M.\ Baake, and H.\ Wagner.
\newblock Ising quantum chain is equivalent to a model of biological evolution.
\newblock {\em Physical Review Letters}, 78(3):559--562, 1997.
\newblock Erratum, Physical Review Letters {\bf 79} (1997), 1782.

\bibitem{BG00}
E.\ Baake and W.\ Gabriel.
\newblock Biological evolution through mutation, selection, and drift: An
  introductory review.
\newblock In D.~Stauffer, editor, {\em Annual Reviews of Computational Physics
  VII}, pages 203--264. World Scientific, Singapore, 2000.

\bibitem{Bur00}
R.\ B{\"{u}}rger.
\newblock {\em The Mathematical Theory of Selection, Recombination, and
  Mutation}.
\newblock Wiley, Chichester, 2000.

\bibitem{Cha90}
B.\ Charlesworth.
\newblock Mutation-selection balance and the evolutionary advantage of sex and
  recombination.
\newblock {\em Genetical Research Cambridge}, 55(3):199--221, 1990.

\bibitem{CCA01}
S.\ Crotty, C.~E.\ Cameron, and R.\ Andino.
\newblock \mbox{RNA} virus error catastrophe: Direct molecular test by using
  ribavirin.
\newblock {\em Proceedings of the National Academy of Sciences},
  98(12):6895--6900, 2001.

\bibitem{DH97}
E.\ Domingo and J.~J.\ Holland.
\newblock \mbox{RNA} virus mutations and fitness for survival.
\newblock {\em Annual Review of Microbiology}, 51:151--178, 1997.

\bibitem{DCCC98}
J.~W. Drake, B.\ Charlesworth, D.\ Charlesworth, and J.~F. Crow.
\newblock Rates of spontaneous mutation.
\newblock {\em Genetics}, 148(4):1667--1686, 1998.

\bibitem{Eig71}
M.\ Eigen.
\newblock Selforganization of matter and the evolution of biological
  macromolecules.
\newblock {\em Naturwissenschaften}, 58(10):465--523, 1971.

\bibitem{GG04}
T.\ Garske and U.\ Grimm.
\newblock A maximum principle for the mutation--selection equilibrium of
  nucleotide sequences.
\newblock {\em Bulletin of Mathematical Biology}, 66(3):397--421, 2004.
\newblock (Preprint physics/0303053).

\bibitem{HRWB02}
J.\ Hermisson, O.\ Redner, H.\ Wagner, and E.\ Baake.
\newblock Mutation selection balance: Ancestry, load, and maximum principle.
\newblock {\em Theoretical Population Biology}, 62:9--46, 2002.

\bibitem{HWB01}
J.\ Hermisson, H.\ Wagner, and M.\ Baake.
\newblock Four-state quantum chain as a model of sequence evolution.
\newblock {\em Journal of Statistical Physics}, 102(1/2):315--343, 2001.

\bibitem{JC69}
T.~H.\ Jukes and C.~R.\ Cantor.
\newblock Evolution of protein molecules.
\newblock In H.~N. Munro, editor, {\em Mammalian Protein Metabolism}, pages
  21--132. Academic Press, New York, 1969.

\bibitem{Kar66}
S.\ Karlin.
\newblock {\em A First Course in Stochastic Processes}.
\newblock Academic Press, New York, 1966.

\bibitem{KS60}
J.~G.\ Kemeny and J.~L.\ Snell.
\newblock {\em Finite Markov Chains}.
\newblock Van Nostrand Reinhold Company, New York, 1960.

\bibitem{Kim80}
M.\ Kimura.
\newblock A simple method for estimating evolutionary rate of base
  substitutions through comparative studies of nucleotide sequences.
\newblock {\em Journal of Molecular Evolution}, 16:111--120, 1980.

\bibitem{Kim81}
M.\ Kimura.
\newblock Estimation of evolutionary distances between homologous nucleotide
  sequences.
\newblock {\em Proceedings of the National Academy of Sciences USA},
  78(1):454--458, 1981.

\bibitem{Kon88}
A.~S.\ Kondrashov.
\newblock Deleterious mutations and the evolution of sexual reproduction.
\newblock {\em Nature}, 336:435--440, 1988.

\bibitem{Leu86}
I.\ Leuth\"ausser.
\newblock An exact correspondence between \mbox{Eigen's} evolution model and a
  two-dimensional \mbox{Ising} system.
\newblock {\em Journal of Chemical Physics}, 84(3):1884--1885, 1986.

\bibitem{Leu87}
I.\ Leuth\"ausser.
\newblock Statistical mechanics of \mbox{Eigen's} evolution model.
\newblock {\em Journal of Statistical Physics}, 48(1/2):343--360, 1987.

\bibitem{MS95}
J.\ {Maynard Smith} and E.\ Szathm{\'a}ry.
\newblock {\em The Major Transitions in Evolution}.
\newblock Freeman, Oxford, 1995.

\bibitem{OK73}
T.\ Ohta and M.\ Kimura.
\newblock A model of mutation appropriate to estimate the number of
  electrophoretically detectable alleles in a finite population.
\newblock {\em Genetical Research}, 22:201--204, 1973.

\bibitem{SOWH96}
D.~L.\ Swofford, G.~J.\ Olsen, P.~J.\ Waddell, and D.~M.\ Hillis.
\newblock Phylogenetic inference.
\newblock In David~M.\ Hillis, Craig Moritz, and Barbara~K.\ Mable, editors,
  {\em Molecular Systematics}, pages 407--514. Sinauer, Sunderland, 1996.

\bibitem{Tar92}
P.\ Tarazona.
\newblock Error thresholds for molecular quasispecies as phase transitions:
  From simple landscapes to spin-glass models.
\newblock {\em Physical Review A}, 45(8):6038--6050, 1992.

\bibitem{vLi82}
J.~H.\ van Lint.
\newblock {\em Introduction to Coding Theory}.
\newblock Springer, Berlin, 1982.

\end{thebibliography}
\end{document}